\documentclass[letterpaper,pre,twocolumn,showpacs,floatfix]{revtex4}
\usepackage{graphicx}
\usepackage{amsmath}
\usepackage{amssymb}

\begin{document}

\title{Swelling of particle-encapsulating random manifolds}

\author{Emir Haleva}
\author{Haim Diamant}
\email{hdiamant@tau.ac.il}
\affiliation{School of Chemistry, Raymond \& Beverly Sackler 
Faculty of Exact Sciences, Tel Aviv University, Tel Aviv 69978, Israel}

\date{August 24, 2008}

\newcommand{\ie}[0] {{\it i.e.},\ }
\newcommand{\eg}[0] {{\it e.g.},\ }

\newcommand{\Fig}[1] {Fig.\ \ref{#1}}
\newcommand{\Eq}[1] {Eq.\ (\ref{#1})}

\newcommand{\KT} {k_{\rm B}T}
\newcommand{\V}{\langle V \rangle}
\newcommand{\Vmax}{V_{\rm max}}
\newcommand{\VmaxTD}{V_{\rm max}}
\newcommand{\lp}{l_{\rm p}}
\newcommand{\pc}{p_{\rm c}}
\newcommand{\Rg}{R_{\rm g}}

\begin{abstract}
  We study the statistical mechanics of a closed random manifold of
  fixed area and fluctuating volume, encapsulating a fixed number of
  noninteracting particles.  Scaling analysis yields a unified
  description of such swollen manifolds, according to which the mean
  volume gradually increases with particle number, following a single
  scaling law. This is markedly different from the swelling under
  fixed pressure difference, where certain models exhibit criticality.
  We thereby indicate when the swelling due to encapsulated particles
  is thermodynamically inequivalent to that caused by fixed pressure.
  The general predictions are supported by Monte Carlo simulations of
  two particle-encapsulating model systems --- a two-dimensional
  self-avoiding ring and a three-dimensional self-avoiding fluid
  vesicle. In the former the particle-induced swelling is
  thermodynamically equivalent to the pressure-induced one whereas in
  the latter it is not.
\end{abstract}


\maketitle

\section{Introduction\label{secIntroduction}}

There has been considerable interest in the past few decades in the
statistical mechanics of membranes and surfaces \cite{Nelson}. This
has been partly motivated by the ubiquity of bilayer membrane vesicles
\cite{Safran1994} in various natural and industrial systems. Since the
lateral size $L$ of such envelopes is much larger than their
thickness, they can be treated to a good approximation as
$(d-1)$-dimensional manifolds, $d$ being the embedding dimension.
Another consequence of the thinness of the membrane is that it resists
stretching much stronger than bending.  Hence, the surface area $A$ of
the membrane is usually assumed fixed. The statistical mechanics of
such a manifold involves an interplay between conformational
fluctuations and bending elasticity, leading to a characteristic
persistence length, $\lp$ \cite{deGennes1982,Lipowsky1991}; over
distances smaller than $\lp$ the manifold is essentially smooth,
whereas beyond it the surface becomes random.  When the manifold is
closed (a vesicle), its smoothness is affected not only by the elastic
persistence length but also by the degree of swelling (\eg
volume-to-area ratio).

The various studies of vesicle thermodynamics can be classified in two
groups according to the volume constraint that they impose (for a
given $A$). One body of works, \eg Refs.\ 
\cite{Seifert1997,Canham1970,Helfrich1973,Seifert1991}, considers the
ensemble of fixed volume $V$. These studies, aimed at actual bilayer
vesicles, assume the low-temperature limit, $\lp>L$, in which the
vesicle is represented by a continuous closed surface in three
dimensions (3D). The various equilibrium shapes are derived as ground
states of the elastic Helfrich Hamiltonian \cite{Helfrich1973},
depending on the dimensionless volume-to-area ratio, $V/A^{3/2}$.
Another body of works treats the ensemble of fixed pressure difference
$p$ across the manifold. The studied systems include Gaussian
\cite{Rudnick1993}, freely jointed \cite{Haleva2006a,Mitra2007}, and
self-avoiding \cite{Fisher1987,Fisher1990,Fisher1990_2} rings in two
dimensions (2D), as well as model fluid vesicles in 3D
\cite{Gompper1992,Gompper1992a,Gompper1992b,Gompper1994,Gompper1997,
Dammann1994,Baumgartner1992}. Most of these works assume the random,
high-temperature limit ($\lp\ll L$), yet the crossover to $\lp>L$ was
addressed as well \cite{Fisher1987,Mitra2007,Gompper1994,Gompper1997}.

As long as equilibrium averages are concerned, the ensembles of fixed
$V$ and fixed $p$ are equivalent, \ie they are related by a smooth,
single-valued (Legendre) transform.  We focus here on another swelling
scenario, where the manifold encapsulates a fixed number $Q$ of
particles while its volume is unconstrained. The interest in such
particle-swollen manifolds is not merely theoretical; most actual
vesicles are immersed in solution and their membrane, over
sufficiently long, experimentally relevant time, is semipermeable,
allowing solvent exchange while keeping the solute trapped inside
\cite{Sackmann1991,Sackmann1999,Peterlin2008}.  Note that the particle
number $Q$ does not {\it a priori} imply a certain osmotic pressure,
because the manifold is free to change its mean volume and, hence, the
mean particle concentration. Nonetheless, since the mean volume and
pressure should monotonously increase with $Q$, one expects to find
equivalence (\ie certain well-behaved transforms) between the
fixed-$Q$ ensemble and the other two.  We have recently demonstrated,
however, that these ensembles are not equivalent for a freely jointed
ring in 2D \cite{Haleva2006b}.  A second-order transition between
crumpled and swollen states, which occurs in the fixed-$p$ ensemble at
a critical pressure $\pc$ \cite{Haleva2006a}, disappears in the
fixed-$Q$ case.  The criticality is avoided as the system selects such
a mean volume that the mean pressure always lies above $\pc$ for any
value of $Q$.  Thus, the regions of phase space covered in the two
ensembles are different.

In the current work we generalize these results, obtaining a unified
description for a $(d-1)$-dimensional random manifold in $d$
dimensions, swollen by either a fixed pressure difference or a fixed
number of trapped particles. We thereby clarify when the swelling
scenarios are thermodynamically equivalent and when they are not.  It
should be borne in mind that, while the current work is focused on
strongly fluctuating, random manifolds ($L\gg\lp$), the bending
rigidity and size of real vesicles place them in the low-temperature,
smooth regime ($L<\lp$) \cite{ft_prl}.

We begin in Sec.\  \ref{secScaling} with a heuristic scaling analysis,
which nevertheless yields the correct qualitative swelling behavior as
found in previously  studied models.  We then proceed  to verify these
general  results  using Monte  Carlo  (MC)  simulations  of two  model
manifolds: a self-avoiding ring  in 2D (Sec.\ \ref{secMC2}), for which
the  two scenarios  are found  to be  equivalent, and  a self-avoiding
fluid vesicle in 3D (Sec.\ \ref{secMC3}), for which they are not.  The
results are analyzed and summarized in Sec.\ \ref{secDiscussion}.

\section{Scaling Analysis\label{secScaling}}

We apply a scaling theory \cite{deGennes,Fisher1990,Pincus1976} to a
closed $(d-1)$-dimensional random manifold, composed of $N$ nodes and
embedded in $d$ dimensions.  In response to perturbation (pressure
difference $p$ or $Q$ noninteracting trapped particles), the manifold
is assumed to be divided into subunits, or blobs, containing $g$ nodes
each. The blobs are defined such that each of them stores a tensile
energy equal to the thermal energy $k_{\rm B}T\equiv 1$
\cite{Pincus1976},
\begin{equation}
  \gamma \xi^{d-1} \sim 1,
  \label{eqGammaBlob}
\end{equation}
where $\gamma$ is the surface tension induced in the manifold due to
the perturbation, and $\xi^{d-1}$ is the projected area of a blob. At
length scales smaller than the blob size $\xi$ the manifold is
unaffected by the perturbation and assumed to obey the power law,
\begin{equation}
  \xi^{d-1} \sim g^\nu,
  \label{eqgBlob}
\end{equation}
where $\nu$ is a swelling exponent characterizing the unperturbed
manifold statistics. At distances larger than $\xi$ the perturbation
stretches the manifold. The total projected area is given by the
number of blobs times the projected area per blob,
\begin{equation}
  R^{d-1} \sim (N/g)\xi^{d-1}.
  \label{eqABlob}
\end{equation}
So far Eqs.\ (\ref{eqGammaBlob})--(\ref{eqABlob}) have been
independent of the nature of perturbation ($p$ or $Q$). The difference
between the two cases enters via the Laplace law, which takes the
following forms in the fixed-$p$ and fixed-$Q$ ensembles,
respectively:
\begin{subequations}
  \begin{eqnarray}
    \label{eqLaplaceP}
    \gamma/R &\sim& p, \\
    \label{eqLaplaceQ}
    \gamma/R &\sim& Q/R^d.
  \end{eqnarray}
\end{subequations}

Solution of Eqs.\ (\ref{eqGammaBlob})--(\ref{eqLaplaceP}) leads to the
following power laws for the fixed-$p$ case:
\begin{subequations}
  \begin{eqnarray}
    \label{eqVP}
    \V &\sim& R^d\sim N^\frac{d}{d-1}(pN^{\frac{1}{d-1}})^\frac{d(1-\nu)}{d\nu-1},\nonumber \\
    \gamma &\sim& (pN^{\frac{1}{d-1}})^\frac{\nu(d-1)}{d\nu -1}.
  \end{eqnarray}
\end{subequations}
(This result, in a different form, has been already obtained in Ref.\
\cite{Gompper1992}.)  Two observations readily follow from
\Eq{eqVP}. First, the characteristic pressure difference, required to
appreciably swell the manifold (\ie to obtain $R\sim N^{1/(d-1)}$),
scales as $p\sim N^{-1/(d-1)}$, regardless of $\nu$. This
characteristic value reflects the interplay between the mechanical
work of inflating an object of volume $\sim N^{d/(d-1)}$, and the
surface entropy of $N$ degrees of freedom, $p N^{d/(d-1)}\sim
N$. Second, in cases where $d\nu=1$ the exponents diverge, \ie the
analysis breaks down, and one expects criticality
\cite{Fisher1990}. Both conclusions are borne out by previously
studied models. Gaussian \cite{Rudnick1993} and freely jointed
\cite{Haleva2006a} rings, having $d=2$ and $\nu=1/2$ (\ie $d\nu=1$)
behave critically at $\pc\sim N^{-1}$, the former swelling to infinite
volume, and the latter undergoing a second-order transition to a
smooth state. By contrast, self-avoiding rings, with $d=2$ and
$\nu=3/4$, swell gradually with $p$
\cite{Fisher1987,Fisher1990,Fisher1990_2}.

Turning to the fixed-$Q$ case, we find from Eqs.\
(\ref{eqGammaBlob})--(\ref{eqLaplaceQ}) the power laws,
\begin{align}
  \V &\sim N^\frac{d}{d-1}(Q/N)^\frac{d(1-\nu)}{d-1}, \notag \\
  \gamma &\sim (Q/N)^\nu. \tag{5b}
  \label{eqVQ}
\end{align}
The corresponding observations in this case are as follows. First,
appreciable swelling occurs for $Q\sim N$, regardless of $\nu$ and
$d$.  Thus, the number of encapsulated particles required to swell the
envelope scales with the area only, rather than the volume. This is a
consequence of considering a vanishing external pressure
\cite{ft_prl}. In such a case the particle entropy ($\sim Q$) has to
compete only with the surface one ($\sim N$). Second, there is no
divergence of exponents in \Eq{eqVQ}, \ie no criticality. Both
conclusions are consistent with findings regarding
particle-encapsulating freely jointed rings in 2D \cite{Haleva2006b}.

The two blob analyses, along with the resulting power laws [Eqs.\
(\ref{eqVP}) and (\ref{eqVQ})], should hold so long as $1<g<N$. This
corresponds to the restrictions, $N^{-d\nu/(d-1)} < p < N^{-1/(d-1)}$,
and $1<Q<N$.  At larger swelling, nonetheless, we expect the manifold
to be smooth, having $\V\sim N^{d/(d-1)}$. According to Laplace's law
this leads to a surface tension $\gamma\sim pN^{1/(d-1)}$ and
$\gamma\sim Q/N$. Combining these large-swelling results with Eqs.\
(\ref{eqVP}) and (\ref{eqVQ}), and provided there is no criticality
($d\nu\neq 1$), we conjecture the following scaling relations,
expected to hold for all values of $p$ and $Q$:
\begin{subequations}
  \begin{eqnarray}
    \label{eqVScalingP} 
    \V &=& N^\frac{d}{d-1} f_p(p N^\frac{1}{d-1}), \nonumber \\
    \gamma &=& h_p(p N^\frac{1}{d-1}), \\
    \label{eqVScalingQ}    
    \V &=& N^\frac{d}{d-1} f_Q(Q/N), \nonumber \\
    \gamma &=& h_Q(Q/N).
  \end{eqnarray}
\end{subequations}
The scaling functions for the mean volume, $f_p$ and $f_Q$, should
cross over from the power laws of Eqs.\ (\ref{eqVP}) and (\ref{eqVQ})
for small arguments to constant values for large arguments.  The
scaling functions for the surface tension, $h_p$ and $h_Q$, are
expected to cross over from the power laws of Eqs.\ (\ref{eqVP}) and
(\ref{eqVQ}) to linear ones.  The validity of Eqs.\ (\ref{eqVQ}) and
(\ref{eqVScalingQ}) has been already proven for a
particle-encapsulating freely jointed ring in 2D \cite{Haleva2006b}.
In addition, the scaling of \Eq{eqVScalingP} has been demonstrated in
the swelling of those rings with increasing $p$ above the critical
point \cite{Haleva2006a}.  We now proceed to check the validity of
Eqs.\ (5)--(6) in two additional model systems.

\section{Self-Avoiding Ring in 2D}\label{secMC2}

We first follow the model and MC scheme presented in Refs.\
\cite{Fisher1987,Fisher1990,Fisher1990_2} for a 2D self-avoiding ring
subject to an inflating pressure difference $p$ . The manifold is
represented by a closed chain of $N$ self-avoiding circles (beads) of
diameter $\sigma=(5/9)l$, linked by tethers of maximum length $l\equiv
1$. In each MC step every bead is moved to a random position within a
square of $(-0.2\sigma,0.2\sigma)^2$ about its former position. These
values of $\sigma$, $l$, and maximum step size ensure that
self-intersection of the ring cannot occur. The move is weighted by
$W=e^{p\Delta V}$, where $\Delta V$ is the difference in (2D) volume
of the ring due to the move, and is accepted provided that (i)
self-avoidance is fulfilled; (ii) tethers do not exceed their maximum
length; and (iii) $W$ exceeds a random number in the range [0,1].
Simulations were performed for $N=50$ to 800.

\begin{figure}[tbh]
  \centering
  \vspace{0.3in} 
  \includegraphics[width=3.4in]{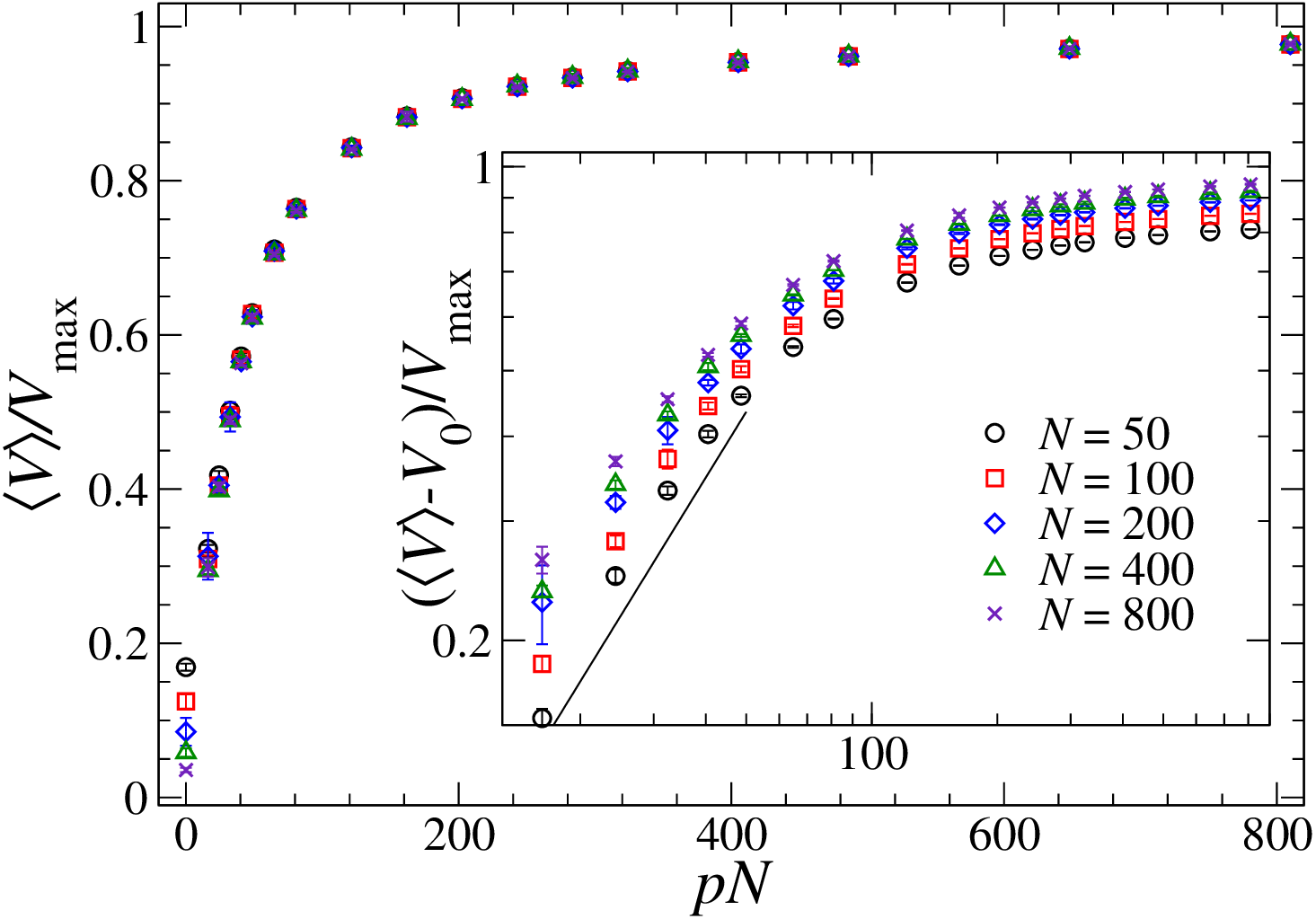}
  \caption{(Color online) Mean volume of 2D self-avoiding rings as a
    function of pressure difference. Data were obtained by MC
    simulations for different ring sizes $N$ and rescaled according to
    \Eq{eqVScalingP}, $\Vmax=N^2/(4\pi)$ being the maximum volume of
    the ring. Inset shows the same data on a log-log scale after the
    mean volume of the unperturbed ring, $V_0\sim N^{3/2}$, has been
    subtracted from $\V$. A solid line of slope 1 is shown for
    reference.}
  \label{figAp}
\end{figure}

The mean volume of the ring as a function of pressure difference is
presented in \Fig{figAp}. The different data sets collapse onto a
single curve once the mean volume is rescaled by the maximum volume of
the ring, $\Vmax= N^2/(4\pi)$, and the pressure by $N^{-1}$, in accord
with \Eq{eqVScalingP}. This scaling law, however, yields a vanishing
mean volume for $p=0$, whereas the unperturbed ring has a finite mean
volume of $V_0\sim N^{2\nu}$, $\nu=3/4$. In the thermodynamic limit
($N\rightarrow\infty$) the correction is negligible, $V_0/\Vmax\sim
N^{-1/2} \rightarrow 0$, but for finite rings the scaling of
\Eq{eqVScalingP} breaks down for sufficiently small $p$, as seen in
\Fig{figAp}. Therefore, to capture the initial linear dependence of
$\V$ on $p$, as predicted by \Eq{eqVP} for $d=2$ and $\nu=3/4$, we
replot in \Fig{figAp} (inset) the data for $\V-V_0$. The initial
increase of $\V-V_0$ with $p$ seems to be consistent with a linear
law, although we cannot claim to have clearly confirmed it.

Next, we turn to particle-encapsulating manifolds by setting $p=0$ and
introducing $Q$ ideal particles at random positions inside the ring.
Hard-core repulsion is introduced between the particles and envelope
beads (but not between the particles themselves), with particle--bead
minimum distance of $\sigma$. The MC step is extended to include
repositioning of each particle within a square of
$(-0.2\sigma,0.2\sigma)^2$ about its former position. This maximum
step size, together with the hard-core repulsion between particles and
envelope beads and maximum tether length, ensure that particles cannot
exit the ring. Rings of $N=50$ to 800 have been simulated, with $Q$
ranging between 0 and $20N$.

\begin{figure}[tbh]
  \centering
  \vspace{0.3in}
  \includegraphics[width=3.4in]{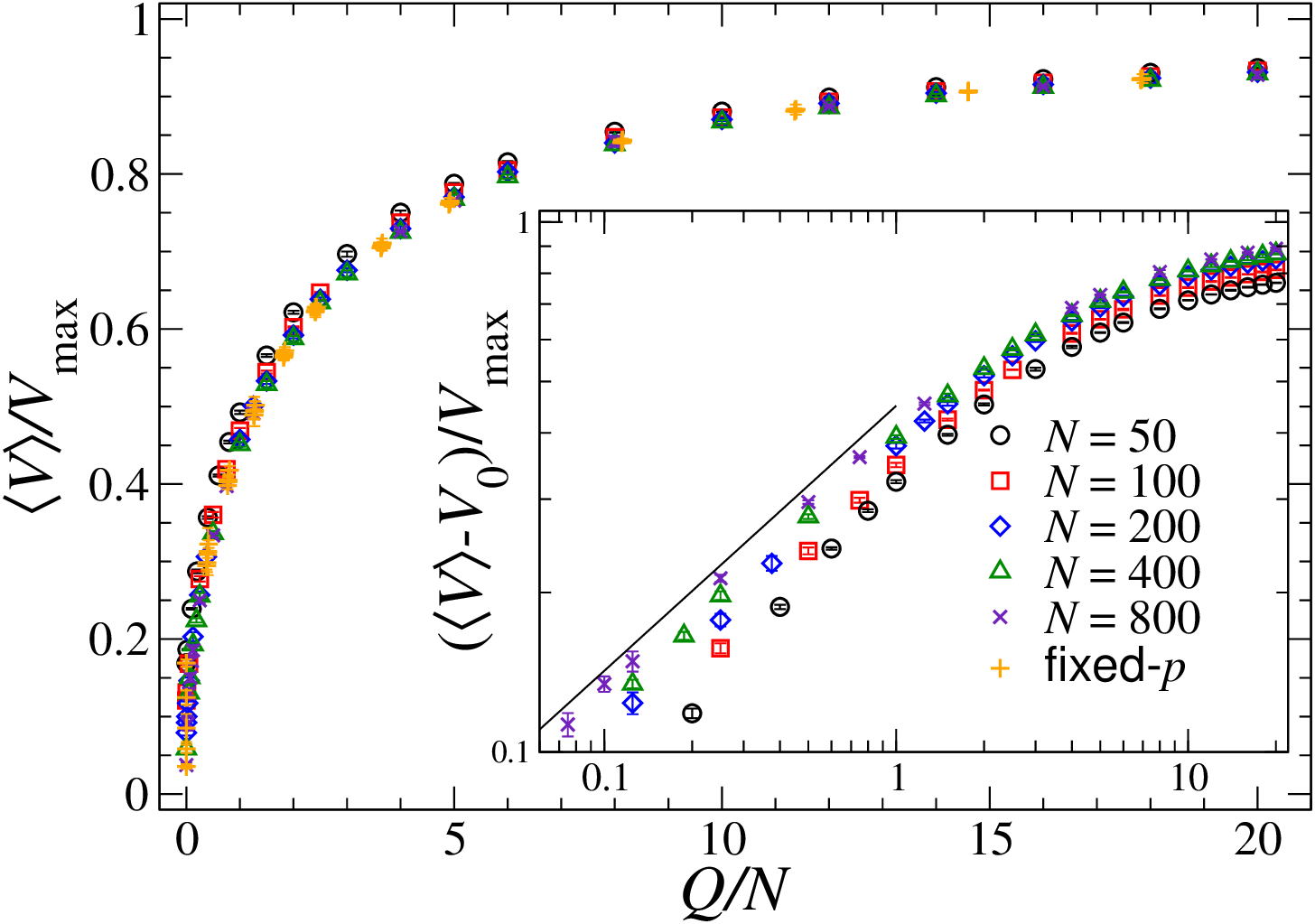}
  \caption{(Color online) Mean volume of 2D self-avoiding rings as a
    function of number of encapsulated particles. Data were obtained
    by MC simulations for different ring sizes $N$ and rescaled
    according to \Eq{eqVScalingQ}, $\Vmax = N^2/(4\pi)$ being the
    maximum volume of the ring. Also plotted are the data points from
    the fixed-$p$ simulation (\Fig{figAp}), whose horizontal
    coordinate is calculated as $p\langle V(p)\rangle/N$. Inset shows
    the data on a log--log scale after the mean volume of the
    unperturbed ring, $V_0\sim N^{3/2}$, has been subtracted from
    $\V$. A solid line of slope 1/2 is shown for reference.}
  \label{figAQ}
\end{figure}

In \Fig{figAQ} we present the mean volume as a function of $Q$. In
agreement with \Eq{eqVScalingQ} all data collapse onto a single curve
when $\V$ is scaled by $N^2$ and $Q$ by $N$. As in the case of fixed
$p$, discussed above, scaling breaks down for very small $Q$, when
$\V$ becomes affected by the finite volume of the unperturbed
state. The power law predicted by \Eq{eqVQ} for 2D self-avoiding rings
($d=2$, $\nu=3/4$), $\V\sim \Vmax (Q/N)^{1/2}$, is nevertheless
verified after subtracting $V_0$ from the mean volume (\Fig{figAQ}
inset).

To demonstrate the equivalence of the fixed-$p$ and fixed-$Q$ scenarios
for this system we transform the data for pressurized rings
(\Fig{figAp}) according to $Q(p)= p\langle V(p)\rangle$ and present
them in \Fig{figAQ} alongside the data for fixed $Q$.  The data sets
of the two scenarios match nicely over the entire ranges of $p$ and
$Q$.


\section{Self-Avoiding Fluid Vesicle in 3D}\label{secMC3}

The second manifold we consider is a discrete model of a fluid
vesicle, which was extensively studied by MC simulations under fixed
pressure difference $p$ \cite{Gompper1992,Gompper1992a,Gompper1992b,
Gompper1994,Gompper1997,Dammann1994}.  The vesicle is represented by a
closed, triangulated, off-lattice network of $N$ nodes (self-avoiding
spheres) of diameter $\sigma=l/\sqrt{2}$, interconnected by a fixed
number of tethers of maximum length $l\equiv 1$.  Membrane fluidity is
mimicked by constantly varying the network connectivity. The MC step
comprises two parts. (i) Each bead is moved randomly within a cube of
$(-0.1\sigma, 0.1\sigma)^3$ about its former position (self-avoidance
permitting).  The move is weighted by a Boltzmann factor of
$e^{p\Delta V}$, where $\Delta V$ is the change in volume caused by
the move.  (ii) $N$ attempts are made to break a randomly chosen
tether, which has formed the common side of two triangles, and rebuild
it between the two other corners of those triangles (provided that the
required tether length does not exceed $l$). The choice of $\sigma$,
$l$, and maximum step size prevents a bead from passing through
another part of the network, making the manifold self-avoiding.

The swelling of this model vesicle as a function of $p$ follows three
regimes \cite{Gompper1992}.  (i) At low pressures the vesicle is in a
collapsed state, having branched-polymer statistics, where the mean
volume and mean-square radius of gyration scale as $\V\sim R^2\sim N$,
with negligible dependence on $p$
\cite{Gompper1992,Baillie1992,Boal1992,Gompper1992c}.  (ii) At a
critical pressure, $p=p^*(N)$, the vesicle undergoes a first-order
transition to a swollen state, whose mean volume gradually increases
with $p$ as $\V\sim p^{0.47}N^{1.73}$ \cite{Gompper1992}.  (iii) At
sufficiently large $p$ the power-law behavior crosses over to
asymptotic swelling toward the maximum volume.

The blob analysis presented in Sec.\ \ref{secScaling} obviously fails
in regime (i) of low swelling, since the volume enclosed in such
collapsed manifolds does not follow the standard relation $\V\sim
R^d$. Instead, we use the fact that the ratio between the
cross-section (frame) area of the manifold and its real surface area
is vanishingly small. Such a manifold has a constant surface tension,
$\gamma\sim 1$ (in units of $\KT/l^2$ \cite{David1991}.  Applying
Laplace's law, $p\sim \gamma/R \sim N^{-1/2}$, we find that the
deflated regime (i) is valid for $p\lesssim N^{-1/2}$, \ie
\begin{equation}
  p\lesssim N^{-1/2}:\quad \V\simeq V_0\sim N.
\label{eqVi}
\end{equation}
In regime (ii) the scaling analysis of Sec.\ \ref{secScaling} holds.
Comparison of the previously obtained power law, $\V\sim
p^{0.47}N^{1.73}$, with \Eq{eqVP} for $d=3$, gives $\nu=0.787$
\cite{Gompper1992}. Our modified scaling analysis [Eqs.\ (\ref{eqVP})
and (\ref{eqVScalingP})] indicates that $p$ is scaled with
$N^{-1/2}$. We note that the power-law dependence of the critical
pressure $p^*$ on $N$ has been controversial \cite{Gompper1997}, with
exponents ranging between $-0.5$ \cite{Gompper1992a} and $-0.69$
\cite{Dammann1994}. The scaling argument for $p>p^*$, together with
\Eq{eqVi} for $p<p^*$, strongly suggest that $p^*\sim N^{-1/2}$
\cite{ft_ring}.

\begin{figure}[tbhp]
  \centering
  \vspace{0.29in}
  \includegraphics[width=3.4in]{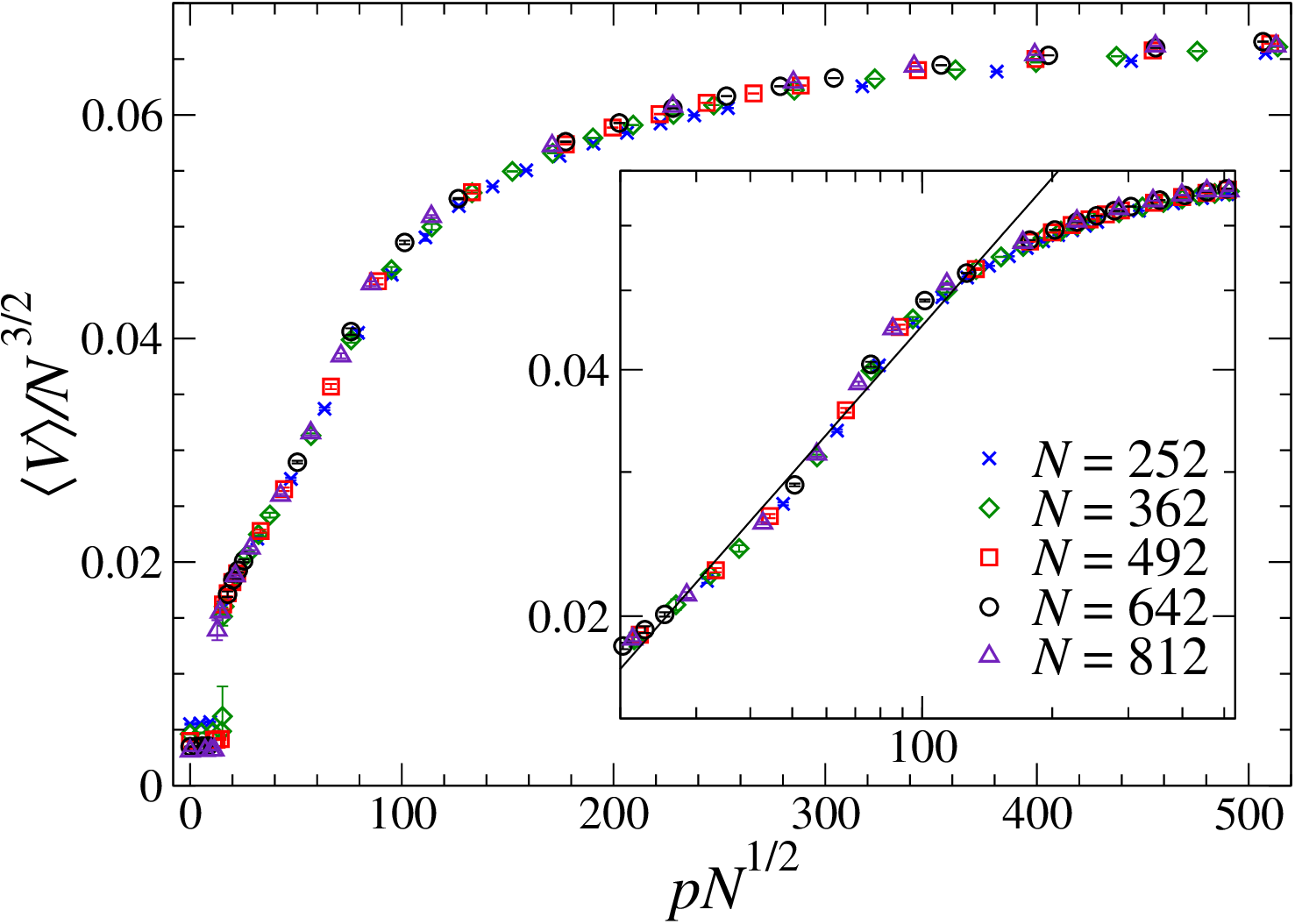}
  \caption{(Color online) Mean volume of 3D self-avoiding fluid
    vesicles as a function of pressure difference, as obtained by MC
    simulations for different vesicle sizes $N$. Data is scaled
    according to \Eq{eqVScalingP}, exhibiting a discontinuous
    transition at $p^*\sim N^{-1/2}$. For $p>p^*$ the data collapse
    onto a single curve. Inset presents the same data for $p>p^*$ on a
    log--log scale. A solid line of slope $0.6$ (corresponding to
    $\nu=0.75$) is shown for reference.}
  \label{figVofP}
\end{figure}

We have repeated the MC simulations for fixed $p$, as presented in
Refs.\ \cite{Gompper1992}, while extending them to larger vesicles and
higher pressure values. The results are shown in \Fig{figVofP},
scaled according to \Eq{eqVScalingP}. The first-order transition at
$p^*\sim N^{-1/2}$ is clearly reproduced, and the predicted scaling
for the entire range of $p>p^*$ is confirmed. The scaling for
$p\gtrsim p^*$ is not inconsistent with the power law of Ref.\
\cite{Gompper1992} and \Eq{eqVP}, having $\nu$ between $0.7$ and $0.8$
(\Fig{figVofP} inset), yet this regime is too narrow to be
clearly resolved.

We now turn to particle-encapsulating vesicles. Repeating the
aforementioned argument for the deflated state of regime (i), $Q/\V
\sim \gamma/R \sim N^{-1/2} $, we find
\begin{equation}
  \langle V\rangle \sim N^{3/2} (Q/N).
  \label{eqVLowSwelling}
\end{equation}
(This linear dependence of $\V$ on $Q$ will be shown in Sec.\
\ref{secDiscussion} to be intimately related to the phase transition
observed as a function of $p$.)  The scaling law of
\Eq{eqVLowSwelling} for the low-swelling regime turns out to be
similar to that of \Eq{eqVQ}. Hence, despite the inadequacy of the
blob analysis in regime (i), we expect the scaling conjecture,
\Eq{eqVScalingQ}, to hold for all values of $Q$ in this model as well.

To check these predictions we modified the MC scheme presented above
by setting $p=0$ and adding $Q$ ideal particles of diameter $\sigma$,
randomly positioned inside the vesicle. The particles do not interact
with each other but have a hard-core repulsion with the network nodes,
keeping them trapped inside the vesicle. The MC step is extended to
include random repositioning of each particle within a cube of
$(-0.1\sigma,0.1\sigma)^3$ about its former position.  Vesicles with
$N$ ranging between 162 and 642 and $Q$ up to $10N$ (for the smallest
vesicle) have been simulated.

\begin{figure}[tbhp]
  \centering
  \vspace{0.29in}
  \includegraphics[width=3.4in]{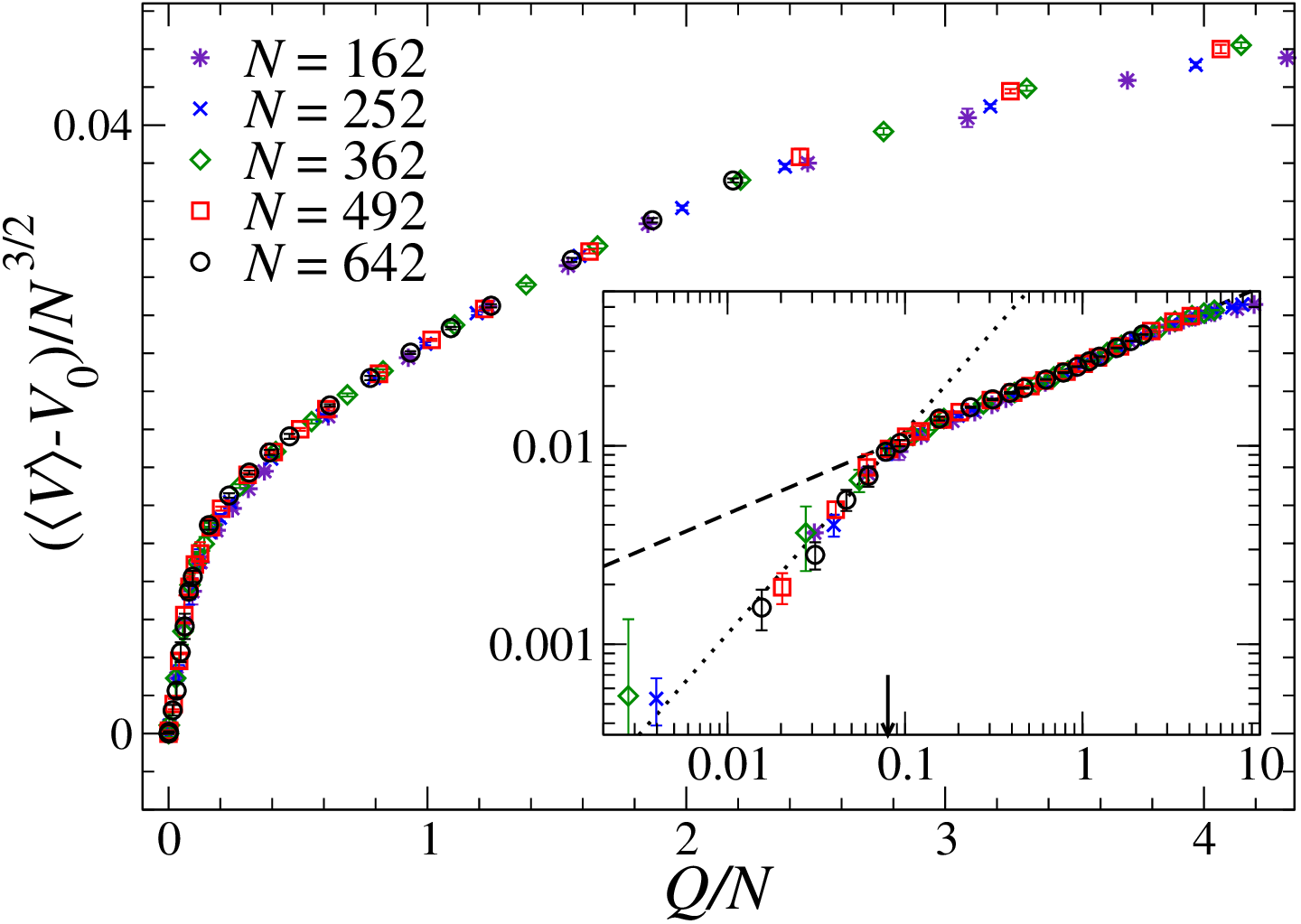}
  \caption{(Color online) Mean volume of 3D self-avoiding fluid
    vesicles as a function of the number of trapped particles, as
    obtained by MC simulations for different vesicle sizes $N$.  Data
    collapse onto a single curve according to \Eq{eqVScalingQ} once
    the volume of the unperturbed vesicle, $V_0\sim N$, is subtracted
    from $\V$. Inset represents the same data on a log-log scale,
    exhibiting a linear regime for $Q\ll N$, $(\V-V_0)/N^{3/2}\sim
    (Q/N)^{1.02(3)}$ (dotted line), followed by a more swollen regime
    with $(\V-V_0)/N^{3/2} \sim (Q/N)^{0.38(3)}$ (dashed line). The
    arrow indicates the crossover between the two regimes at $Q\simeq
    0.08N$.}
  \label{figV2Q}
\end{figure}

Results for the mean volume as a function of $Q$ for various vesicle
sizes are shown in \Fig{figV2Q}. Once the volume $V_0$ of the
unperturbed (branched) state [\Eq{eqVi}], which is inaccessible to
particles due to their excluded-volume interaction with the manifold,
is subtracted from $\V$, the data collapse onto a single curve
according to \Eq{eqVScalingQ}. Two power-law regimes are seen in
\Fig{figV2Q} (inset).  At low swelling $\V$ increases linearly with
$Q$, in agreement with \Eq{eqVLowSwelling} \cite{ft_wrong}. At about
$Q\simeq 0.08N$ the swelling crosses over to a different power law
which, when fitted to \Eq{eqVQ}, yields $\nu=0.75(2)$.  This value is
close to that found in the fixed-$p$ simulations, $\nu=0.787$
\cite{Gompper1992,ft_Kantor}. For larger values of $Q$ this power-law
regime should cross over to asymptotic saturation toward the maximum
volume. Because of computer limitations we could sample only the
lowest edge of this regime (\Fig{figV2Q} inset).

\begin{figure}[tbh]
  \centering
  \vspace{0.1in}
  {
    \includegraphics[width=3.4in]{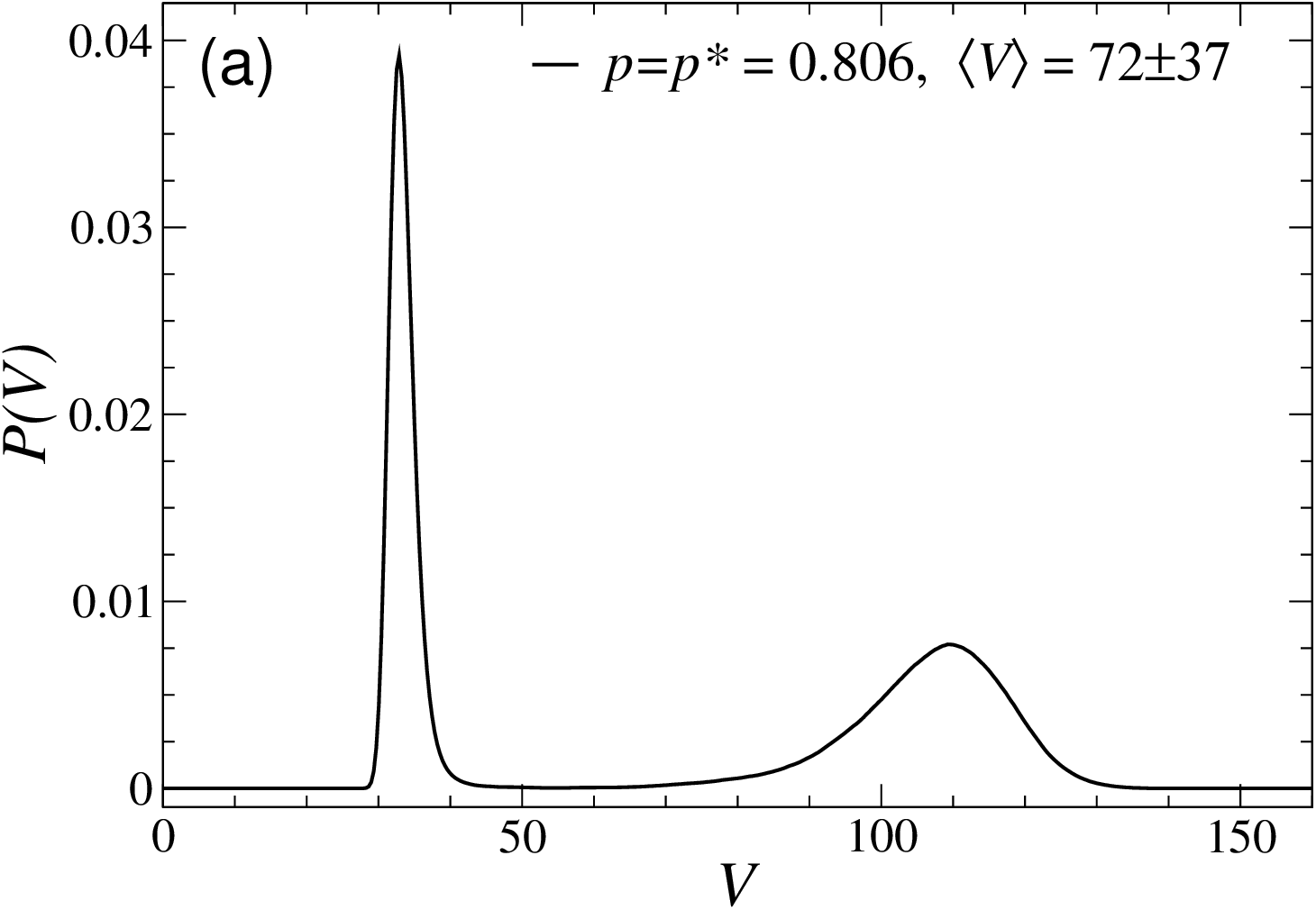}
  }
  \vspace{0.27in}
  {
    \includegraphics[width=3.4in]{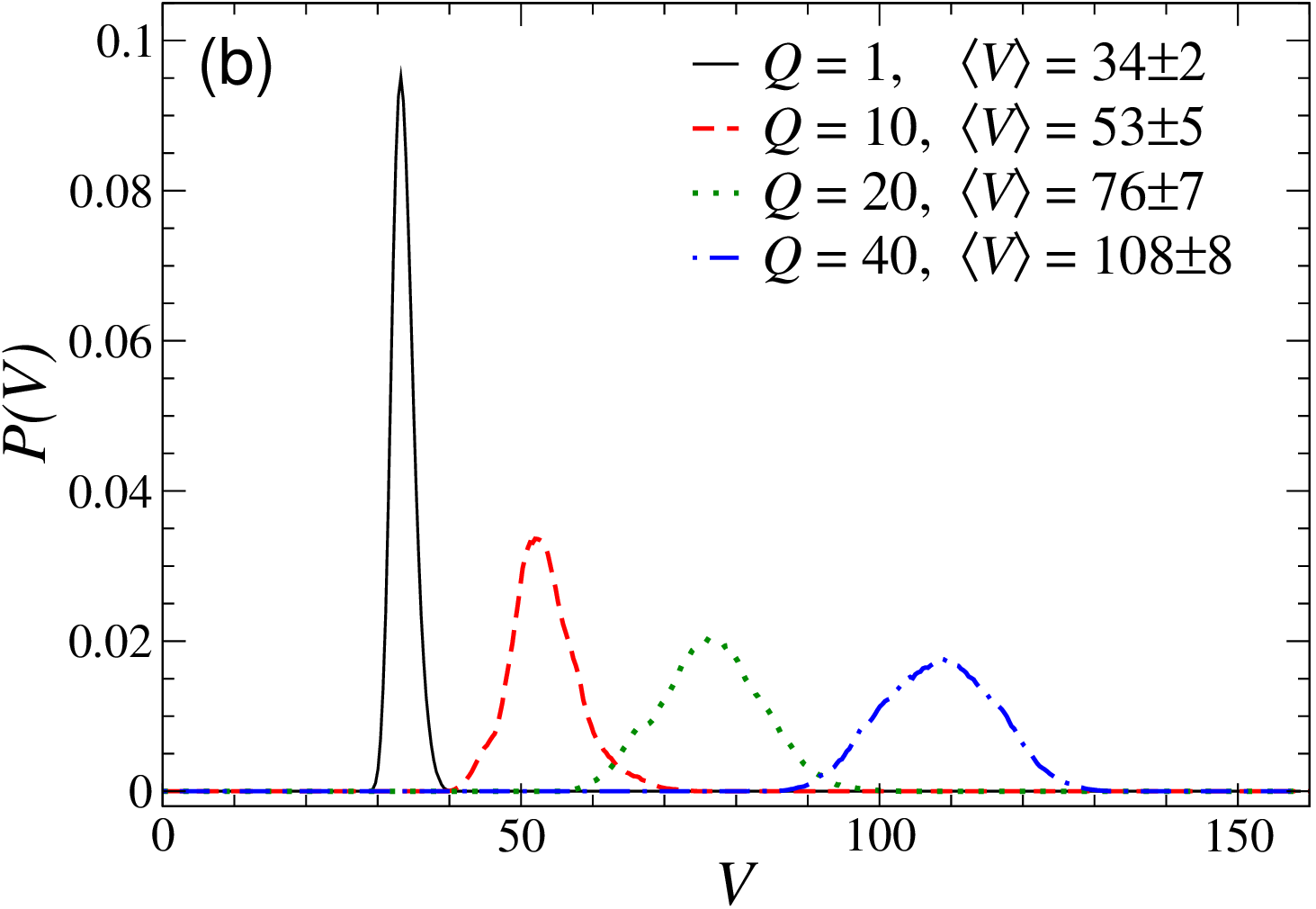}
  }
  \caption{(Color online) Volume distribution functions of 362-node 3D
    vesicles. (a) For fixed pressure difference, at the transition
    point, the distribution is bimodal. (b) Under fixed number of
    encapsulated particles the distribution is always unimodal.  }
  \label{figPoV}
\end{figure}

Unlike the case of fixed $p$, the vesicle gradually swells with $Q$,
exhibiting no phase transition. To further verify the absence of a
first-order transition we have measured the probability distribution
function of the volume, $P(V)$, as a function of $Q$. Whereas under
fixed $p$, at $p=p^*$, one finds a bimodal distribution [Ref.\
\cite{Gompper1992} and \Fig{figPoV}(a)], \ie coexistence of collapsed
and swollen states, for particle-encapsulating vesicles we obtain
unimodal distributions for all values of $Q$ [\Fig{figPoV}(b)].

\begin{figure}[tbh]
  \centering
  \vspace{0.3in}
  \includegraphics[width=3.4in]{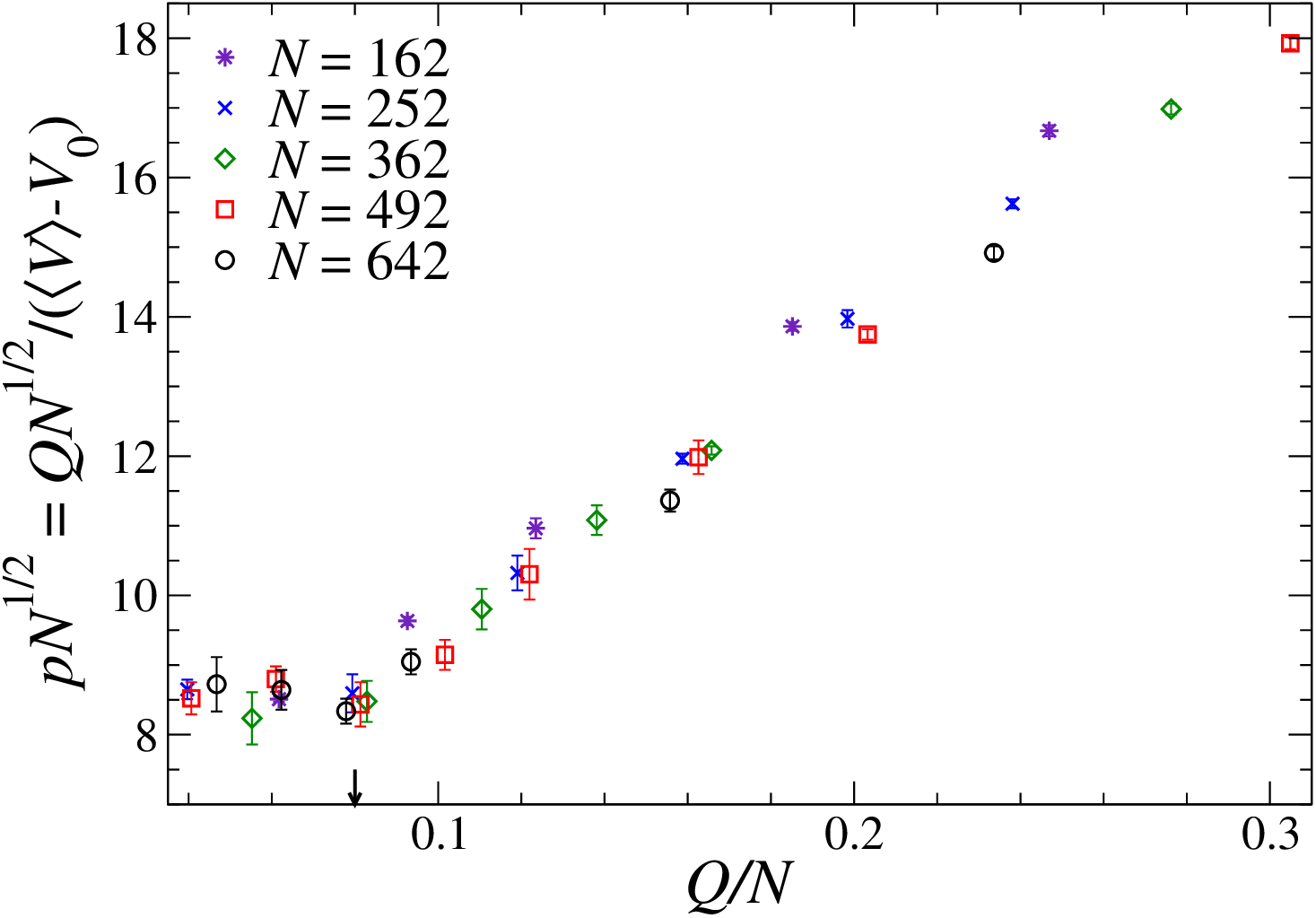}
  \caption{(Color online) Effective pressure of encapsulated
    particles as a function of particle number, as obtained by MC
    simulations for various vesicle sizes $N$. Data collapse according
    to \Eq{eqPsi}. Note the finite effective pressure at small $Q$.
    The arrow indicates the crossover between the two swelling regimes
    at $Q\simeq 0.08N$.}
  \label{figC}
\end{figure}

Finally, let us consider the effective pressure exerted by the
encapsulated ideal particles, $p=Q/\V$. From \Eq{eqVScalingQ} we have
\begin{equation}
  p = N^{-1/2}\psi(Q/N),\quad \psi(x)=x/f_Q(x). 
\label{eqPsi}
\end{equation}
In the low-swelling regime we have found a linear behavior, $f_Q(x\ll
1)\sim x$, [\Eq{eqVLowSwelling} and \Fig{figV2Q} inset]. Thus, $\psi(x
\ll 1)=\mbox{const}$, \ie the effective pressure does not change with
$Q$ throughout this regime. Figure \ref{figC} demonstrates the data
collapse according to \Eq{eqPsi}, as well as the finite, constant
pressure $p_{\rm min}$ at low swelling even for the smallest values of
$Q$. (In calculating the concentration and pressure from the
simulations we have considered the particle-accessible volume,
$\V-V_0$.)  One expects $p_{\rm min}$ to coincide with the transition
value under fixed pressure, $p^*$. [Compare also \Fig{figPoV}(a),
  plotted for $p=p^*$, with \Fig{figPoV}(b), where the effective
  pressure is essentially fixed at $p_{\rm min}$ for all curves.]  We
find, however, $p^*\simeq 1.8p_{\rm min}$. This discrepancy may stem
from the interaction of the particles with the vesicle, making them
deviate from the ideal-gas behavior, particularly in the deflated
state.
 
\section{Discussion}\label{secDiscussion}

The scaling analysis presented in Sec.\ \ref{secScaling} yields a
unified account of the swelling of random manifolds with increasing
pressure difference or number of encapsulated particles. The validity
of this description has been demonstrated for several model systems in
Secs.\ \ref{secMC2} and \ref{secMC3} and in Ref.\ \cite{Haleva2006b}.
Similar scaling analyses for the case of fixed $p$ were previously
presented in Refs.\ \cite{Fisher1990,Gompper1992}. Those analyses and
ours coincide in the power-law regime, \Eq{eqVP}. However, while the
previous analyses are focused on the weak-swelling regime and
constructed to include the random, unperturbed state of the manifold,
the one presented here is aimed at encompassing the high-swelling
behavior. Thus, on the one hand, our scaling relations, Eqs.\
(\ref{eqVScalingP}) and (\ref{eqVScalingQ}), cannot account for the
unperturbed state and give a vanishing mean volume in the limit of
vanishing perturbation. The range of $p$ (or $Q$) where this
deficiency is relevant, nonetheless, vanishes in the thermodynamic
limit \cite{ft_deficiency}. On the other hand, whereas the previous
analyses assumed that scaling broke down at sufficiently large
swelling \cite{Fisher1990,Gompper1992}, we have claimed that Eqs.\
(\ref{eqVScalingP}) and (\ref{eqVScalingQ}) should hold for the entire
range of $p$ or $Q$.  Although there is {\it a priori} no reason why
the scaling behavior should have this broad range, we analytically
proved the conjecture for freely jointed rings in 2D at fixed
$p>p_{\rm c}$ \cite{Haleva2006a} or fixed $Q$
\cite{Haleva2006b}. In the current work we have provided further
numerical support of the scaling conjecture in several additional
systems --- 2D self-avoiding rings at fixed $p$ or fixed $Q$, and 3D
fluid vesicles at fixed $p>p^*$ or fixed $Q$. Hence, provided that the
swelling exhibits no criticality, the scaling relations Eqs.\ 
(\ref{eqVScalingP}) and (\ref{eqVScalingQ}) seem to be applicable in a
broad range of systems.

The analysis presented here for the swelling with $Q$ yields new
scaling relations, which have been confirmed in all studied
systems. The different behavior of particle-encapsulating manifolds
lies in the response of manifold volume to changes in the number of
encapsulated particles. The resulting particle concentration and
effective pressure depend on this response and, therefore, may have a
nontrivial dependence on $Q$. In certain cases this may lead to
thermodynamic inequivalence of the fixed-$p$ and fixed-$Q$ ensembles.
Equivalence breaks down when the two ensembles are no longer related
by a one-to-one smooth transform. In the two examples where
inequivalence has been demonstrated --- 2D freely jointed rings
\cite{Haleva2006b} and 3D fluid vesicles (Sec.\ \ref{secMC3}) --- both
conditions of smoothness and single-valuedness are violated: (i) a
criticality in the fixed-$p$ ensemble makes the transform
$Q(p)=p\langle V(p)\rangle$ nonanalytic; and (ii) the effective
pressure in the fixed-$Q$ ensemble is bounded from below by a finite
value, \ie states of low pressure are inaccessible (cf.\
\Fig{figC}). We now show that this combination of criticality under
fixed $p$ and inaccessible states for fixed $Q$ is not a coincidence.

Let us consider a general power-law response to particle number,
$\V\sim Q^\alpha$.  Transforming to the fixed-$p$ ensemble, we get
$p(Q)=Q/\V\sim Q^{1-\alpha}$ and $\V \sim p^{\alpha/(1-\alpha)}$.
Several observations follow from these relations.  First,
thermodynamic stability dictates that $\V$ increase with $p$, \ie
$\alpha\leq 1$. We are left with two different cases. (i) If
$\alpha<1$, there is no criticality and arbitrarily small values of
$Q$ will correspond to arbitrarily small values of $p$. Hence, in this
case there is equivalence. (ii) If $\alpha=1$, we expect both
criticality under fixed $p$ and inaccessibility of small-pressure
states at fixed $Q$, \ie inequivalence of the two swelling scenarios.
For maximum $\alpha$ the manifold volume is maximally susceptible to
changes in $Q$ (linear in $Q$), to the extent that the concentration
and pressure do not change with $Q$ (cf.\ \Fig{figC}).  Thus,
criticality and inequivalent phase spaces come hand in hand.
In a standard case, where the blob analysis of Sec.\ \ref{secScaling}
holds, we get from \Eq{eqVQ} $\alpha=d(1-\nu)/(d-1)$, and the
condition $\alpha\leq 1$ is equivalent to $d\nu\geq 1$. In addition,
one has a geometrical lower bound for the swelling exponent, which
cannot be smaller than that of a folded, compact manifold,
$\nu\geq(d-1)/d$. This leads to the restriction $\alpha\leq 1/(d-1)$,
which is consistent with, and stricter than, the thermodynamic one,
$\alpha\leq 1$. Hence, we conclude that for most systems, which obey
the analysis of Sec.\ \ref{secScaling}, case (ii) above, involving
criticality and inequivalence, can occur only in 2D, \ie for $d=2$ and
$\nu=1/2$ \cite{Haleva2006a,Haleva2006b}.

All of these general conclusions are supported by specific examples.
A 2D self-avoiding ring is an example of case (i) above. It obeys the
scaling analysis of Sec.\ \ref{secScaling} with $d=2$, $\nu=3/4$, \ie
$\alpha=1/2$. (The value of $\alpha$ has been confirmed by
simulations; see \Fig{figAQ}.) This system exhibits no criticality
under fixed $p$, and the two ensembles have been found equivalent
(\Fig{figAQ}). The more interesting case (ii) has been encountered in
three systems. Two examples are provided by Gaussian and freely
jointed rings in 2D \cite{Haleva2006b}. For both examples the blob
analysis holds, and $d=2,\nu=1/2$ (\ie $d\nu=1$).  The third example
of the anomalous case (ii) is a 3D fluid vesicle (Sec.\ \ref{secMC3}),
for which the blob analysis of Sec.\ \ref{secScaling} fails, yet a
linear dependence of $\V$ on $Q$ has been found [\Eq{eqVLowSwelling}
and \Fig{figV2Q}]. Indeed, under fixed $p$
\cite{Rudnick1993,Haleva2006a,Gompper1992} all three examples exhibit
phase transitions.

This host of examples leads to the expectation that the picture
described here, including the scaling relations and possible phase
transitions, should hold for any random manifold swollen by either a
pressure difference or encapsulated particles.  In cases where the
blob analysis of Sec.\ \ref{secScaling} is valid, one needs to know
merely the dimensionality $d$ and the statistics of the unperturbed
manifold ($\nu$) to predict the qualitative swelling behavior. In
other, exceptional cases (\eg the 3D fluid vesicle of Sec.\
\ref{secMC3}) it suffices to know the response of the unperturbed
manifold to a small number of encapsulated particles (\ie $\alpha$).

The thermodynamic inequivalence between the fixed-$p$ and fixed-$Q$
scenarios, reported above for certain systems, also implies
inequivalence between the canonical and grand-canonical ensembles in
those systems. This is because fixing the chemical potential $\mu$ of
the encapsulated particles inevitably fixes also the mean pressure $p$
that they exert on the manifold, as these two intensive variables are
related via the particles' equation of state. (For example, for ideal
particles $\mu=\ln p$.) Once again, because of the unconstrained
volume, the system cannot attain arbitrarily small concentrations as
$Q$ is decreased, and, therefore, the full range of $\mu$ is not
covered. The inequivalence of the fixed-$Q$ and fixed-$\mu$ ensembles
has been directly demonstrated for freely jointed 2D rings
\cite{Haleva2006b}.

In the current work we have not explicitly considered the bending
rigidity of the vesicle. Such a bending-free description is valid in
two limits: (i) At sufficiently strong swelling the fluctuations of
any vesicle are governed by surface tension rather than bending
rigidity. (ii) If the manifold is sufficiently large ($L\gg\lp$),
bending rigidity merely renormalizes the molecular length $a$ to $\lp$
and the number of surface degrees of freedom $N$ to $Na^2/\lp^2$. It
is this random, strongly fluctuating case which has been the focus of
the current work.  On the one hand, due to their bending rigidity
($\kappa\sim 10$ $k_{\rm B}T$) and size ($0.1$--$10$ $\mu$m), real
bilayer vesicles are smooth and do not satisfy limit (ii). On the
other hand, some of our most significant results (\eg the
inaccessibility of low-pressure states) concern weak swelling, outside
limit (i), where the bending rigidity of real vesicles plays an
important role. Thus, the direct relevance of the current work to real
bilayer vesicles is limited.
Yet, overall, this work and the specific examples associated with it
highlight the qualitative differences which may emerge between
pressurized manifolds and particle-encapsulating ones. Indeed, the
different behavior of particle-encapsulating vesicles is manifest also
in realistic scenarios involving smooth membranes, \eg highly swollen
bilayer vesicles in solution \cite{ft_prl}.

\begin{acknowledgments}
  We thank D.\ Harries for helpful discussions. Acknowledgment is made
  to the Donors of the American Chemical Society Petroleum Research
  Fund for support of this research (Grant no.\ 46748-AC6).
\end{acknowledgments}

\bibliographystyle{unsrt}

\begin{thebibliography}{}
\bibitem{ft_wrong} Artificially attempting to extract $\nu$ for the
  low-swelling regime by fitting these results to \Eq{eqVQ}, one gets
  the unphysical value of $\nu=1/3$, which is smaller than the lower
  bound of $2/3$ set by a fully collapsed, compact manifold. This
  highlights the necessity of a different scaling analysis in this
  regime, as presented in Sec.\ \ref{secMC3}.
\bibitem{ft_deficiency} The relevant range in the fixed-$p$ ensemble
  is $pN^{1/(d-1)}<N^{(1-d\nu)/(d-1)}$, which vanishes as
  $N\rightarrow \infty$ for $d\nu>1$, \ie so long as there is no
  criticality. In the fixed-$Q$ case the range is $Q/N<1/N$, which is
  irrelevant regardless of $\nu$.
\end{thebibliography}

\end{document}